\documentclass[nofootinbib,prd,twocolumn,showpacs,showkeys,preprintnumbers]{revtex4}
\usepackage{hyperref,amssymb,amsmath,mathrsfs,bm,graphicx}
\begin{document}
\title {Irreversibility and gravitational radiation: A proof of  Bondi's conjecture}
\author{L. Herrera}
\email{lherrera@usal.es}
\affiliation{Instituto Universitario de F\'isica
Fundamental y Matem\'aticas, Universidad de Salamanca, Salamanca 37007, Spain}
\author{A. Di Prisco}
\email{alicia.diprisco@ucv.ve}
\affiliation{Escuela de F\'\i sica, Facultad de Ciencias, Universidad Central de Venezuela, Caracas 1050, Venezuela}
\author{J. Ospino}
\email{j.ospino@usal.es}
\affiliation{Departamento de Matem\'atica Aplicada and Instituto Universitario de F\'isica
Fundamental y Matem\'aticas, Universidad de Salamanca, Salamanca 37007, Spain}
\date{\today}
\begin{abstract}
It is shown that the evolution of an axially and reflection symmetric fluid distribution, satisfying the Tolman condition for thermal equilibrium, is not accompanied by the emission of gravitational radiation. This result, which was conjectured by Bondi many years ago, expresses the irreversibility associated to the emission of gravitational waves. The observational consequences emerging from this result are commented. The resulting models are not only non--dissipative and vorticity free, but also shear--free and geodesic, furthermore all their complexity factors vanish.

 \end{abstract}
\date{\today}
\maketitle

\section{Introduction}
In his seminal paper on gravitational radiation \cite{7} Bondi wrote (section 6):
``....If the distinction between radiative and non-radiative motions is locally significant then the clearest self-consistent distinction appears to be between cases where the equations of state do not involve the time explicitly and are time reversible (no dissipation), and others.......''

In other words, the irreversibility of the  process of emission of gravitational waves, must be reflected in the equation of state of the source through an entropy increasing  (dissipative) factor. 

The rationale supporting this conjecture is very clear: radiation is an irreversible process, this fact emerges at once  if  absorption is taken into account and/or Sommerfeld type conditions, which eliminate inward traveling waves, are imposed. Therefore,  it is obvious that an entropy generator factor should be present in the description of the source.  

However, since the Bondi's work deals exclusively with the space--time outside the source (more so, far from the source), the above mentioned relationship between gravitational  radiation and dissipative processes within the source, remained so far  a conjecture (a very reasonable one though).

It is the purpose of this work to provide a definitive proof of the Bondi's conjecture.

For doing that we shall resort to a general formalism to describe the evolution of dissipative axially and reflection symmetric fluid distribution presented in \cite{1}, based in 
 the $1+3$ formalism developed in \cite{21cil, n1, 22cil, nin}.

Our proof develops in two steps. We shall first prove that assuming the Tolman condition  \cite{Tolman}  to be satisfied (implying the absence of dissipative flux), the fluid is necessarily  vorticity free. Next, using this last condition and the absence of dissipation we shall prove that the the magnetic part of the Weyl tensor vanishes. This last result closes the proof of the Bondi's conjecture, since it implies the vanishing of the super--Poynting vector.  Indeed, in the theory of  the super--Poynting vector, a state of gravitational radiation is associated to a  non--vanishing component of the latter (see \cite{11p, 12p, 14p}). This in  turn is in agreement with the established link between the super--Poynting vector and the news functions \cite{5p}, in the context of the Bondi--Sachs approach \cite{7, 8}. 

Besides we shall see that the fluid is necessarily shear--free, geodesic and all their complexity factors vanish.

In the next section we shall briefly summarize the main equations required for our proof. Then we shall proceed with the proof following the steps outlined  before. Finally we discuss about the physical relevance of our results. Some basic definitions and intermediate  formulae are given in the Appendix.

\section{The metric and the source: basic equations and notation}
As mentioned before, we shall resort to the general approach fully deployed in \cite{1} in order to achieve our goal. In this section we shall present very briefly the most general properties of the space--time under consideration and the matter content of the source. The reader is referred to  	\cite{1} and the Appendix, for any specific detail of calculation.

We shall consider  axially (and reflection) symmetric sources. For such a system the most general line element may be written in ``Weyl spherical coordinates'' as:

\begin{equation}
ds^2=-A^2 dt^2 + B^2 \left(dr^2
+r^2d\theta^2\right)+C^2d\phi^2+2Gd\theta dt, \label{1b}
\end{equation}
where $A, B, C, G$ are  functions of $t$, $r$ and $\theta$, of class $C^\omega$, with  $A, B, C$ positive defined.  We number the coordinates $x^0=t, x^1=r, x^2= \theta, x^3=\phi$. At this point it is important to stress that due to the reflection symmetry (no $dtd\phi$ terms in (\ref{1b})), rotations around the symmetry axe are excluded, and vorticity is associated with motion along the $\theta$-direction.

The inverse components of the metric are given by 
\begin{eqnarray}
g^{\alpha \beta}= \left(\begin{array}{cccc}
- \frac{B^2 r^2}{A^2B^2r^2+G^2}  &  0  &   \frac{G}{A^2B^2r^2+G^2}    &   0    \\0 &  \frac{1}{B^2} &
0     &   0    \\ \frac{G}{A^2B^2r^2+G^2}       &  0 &  \frac{A^2}{A^2B^2r^2+G^2} &
0      \\0 &
0       &  0   &  \frac{1}{C^2} \end{array}
\right). \nonumber \\
\label{gar}
\end{eqnarray}
We shall assume that  our source is filled with an anisotropic and dissipative fluid. 
The energy momentum tensor may be written in the ``canonical'' form, as 
\begin{equation}
{T}_{\alpha\beta}= (\mu+P) V_\alpha V_\beta+P g _{\alpha \beta} +\Pi_{\alpha \beta}+q_\alpha V_\beta+q_\beta V_\alpha.
\label{6bis}
\end{equation}

The above is the canonical, algebraic decomposition of a second order symmetric tensor with respect to unit timelike vector, which has the standard physical meaning where $T_{\alpha \beta}$ is the energy-momentum tensor describing some energy distribution, and $V^\mu$ the four-velocity assigned by certain observer. In our case we are considering an Eckart frame  where fluid elements are at rest.

With the above definitions it is clear that $\mu$ is the energy
density (the eigenvalue of $T_{\alpha\beta}$ for eigenvector $V^\alpha$), $q_\alpha$ is the  heat flux, whereas  $P$ is the isotropic pressure, and $\Pi_{\alpha \beta}$ is the anisotropic tensor.

Since we choose the fluid to be comoving in our coordinates, then
\begin{equation}
V^\alpha =\left(\frac{1}{A},0,0,0\right), \quad  V_\alpha=\left(-A,0,\frac{G}{A},0\right).
\label{m1}
\end{equation}
Next, let us  introduce the unit, spacelike vectors $\bold K, \bold L$, $\bold S$, with components
\begin{equation}
K_\alpha=\left(0,B,0,0\right), \quad K^\alpha=\left(0,\frac{1}{B},0,0\right),
\label{7}
\end{equation}
\begin{equation}
L^\alpha=\left(\frac{G}{A\sqrt{A^2B^2r^2+G^2}},0,\frac{A}{\sqrt{A^2B^2r^2+G^2}},0\right),
\label{7c}
\end{equation}
\begin{equation}
L_\alpha=\left(0,0,\frac{\sqrt{A^2B^2r^2+G^2}}{A},0\right),
\label{7d}
\end{equation}

\begin{equation}
S_\alpha=\left(0,0,0,C\right), \quad S^\alpha=\left(0,0,0,\frac{1}{C}\right),
\label{3n}
\end{equation}

satisfying  the following relations:
\begin{equation}
V_{\alpha} V^{\alpha}=-K^{\alpha} K_{\alpha}=-L^{\alpha} L_{\alpha}=-S^{\alpha} S_{\alpha}=-1,
\label{4n}
\end{equation}
\begin{equation}
V_{\alpha} K^{\alpha}=V^{\alpha} L_{\alpha}=V^{\alpha} S_{\alpha}=K^{\alpha} L_{\alpha}=K^{\alpha} S_{\alpha}=S^{\alpha} L_{\alpha}=0.
\label{5n}
\end{equation}
The unitary vectors $V^\alpha, L^\alpha, S^\alpha, K^\alpha$ form a canonical  orthonormal tetrad (say  $e^{(a)}_\alpha$), such that  $$e^{(0)}_\alpha=V_\alpha,\quad e^{(1)}_\alpha=K_\alpha,\quad
e^{(2)}_\alpha=L_\alpha,\quad e^{(3)}_\alpha=S_\alpha,$$ with $a=0,\,1,\,2,\,3$ (latin indices labeling different vectors of the tetrad). The  dual vector tetrad $e_{(a)}^\alpha$  is easily computed from the condition 

$$ \eta_{(a)(b)}= g_{\alpha\beta} e_{(a)}^\alpha e_{(b)}^\beta.$$

We  shall express all kinematical and physical variables, as well as the equations relating them, in terms of their tetrad components. These expressions are explicitly deployed in the Appendix.

\section{Proving that no--dissipation implies no gravitational radiation}
In order to ensure the absence of dissipation we have to impose the  Tolman conditions for thermodynamic equilibrium  \cite{Tolman}. Such conditions emerge from the fact that, according to special relativity, all forms of energy have inertia, and therefore this should also apply to heat. Then, because of the equivalence principle, there should also be some weight associated to heat, and one should expect that thermal energy tends to displace to regions of lower gravitational potential. This in turn implies that the condition of thermal equilibrium in the presence of a gravitational field must change with respect to its form in the absence of gravity. Thus, a temperature gradient is necessary in thermal equilibrium in order to prevent the flow of heat from regions of higher to lower gravitational potential. Tolman deduced such conditions without any reference to any specific transport equation, however, as expected, for any consistent transport equation, the absence of dissipation should lead to Tolman conditions.

Thus,  for example, in the 
M\"{u}ller-Israel-Stewart second
order phenomenological theory for dissipative fluids \cite{18, 19, 20, 21}), the transport eqution reads reads

\begin{equation}
\tau h^\mu_\nu q^\nu _{;\beta}V^\beta +q^\mu=-\kappa
h^{\mu\nu}(T_{,\nu}+T a_\nu)-\frac{1}{2}\kappa T^2\left
(\frac{\tau V^\alpha}{\kappa T^2}\right )_{;\alpha}q^\mu,\label{qT}
\end{equation}

\noindent where $\tau$, $\kappa$, $T$ denote the relaxation time,
the thermal conductivity and the temperature, respectively.

From (\ref{qT}) we see that the absence of dissipative flux implies at once
\begin{equation}
h^{\mu\nu}(T_{,\nu}+T a_\nu)=0,
\label{tc1}
\end{equation}
which are the Tolman conditions.

We have now all the ingredients required for our proof. Some relevant equations are written down in the Appendix.

We shall assume that the system is in thermodynamic equilibrium, implying that the Tolman conditions  (\ref{tc1}) are satisfied, i.e.
\begin{equation}
a_\mu=-h^\nu_\mu \Gamma_{,v} \qquad \Gamma\equiv \ln T.
\label{ta1}
\end{equation}

From the above equation it follows that
\begin{equation}
a_1=-\Gamma^\prime, \qquad a_2=-\frac{G \dot \Gamma}
{A^2}-\Gamma_{,\theta}.
\label{ta2}
\end{equation}

Using (\ref{ta2}) in (\ref{T7}) produces

\begin{equation}
K^{[\mu}L^{\nu]}a_{\mu;\nu}= V^\mu\Gamma_{,\mu} \Omega,
\label{T7nd}
\end{equation}

which combined with  (\ref{esc51KL}) produces

\begin{equation}
\Omega _{,\delta}V^\delta +\frac{1}{3}(2\Theta+\sigma _I+\sigma _{II}+3V^\mu \Gamma_{,\mu})\Omega =0.\label{ta3}
\end{equation}

The consequences derived from the above equation are far reaching. Indeed, if we assume that the system is initially static (at $t=0$ say), and assume that it starts to evolve afterward, keeping the thermodynamic equilibrium, then the evolving fluid would be vorticity--free.
This result is in full agreement with earlier works indicating that vorticity generation is sourced by entropy gradients 
\cite{Croco}--\cite{75}. At the same time this result reinforces further the Bondi's conjecture about the absence of radiation for non--dissipative systems, if we recall the radiation--vorticity link discussed in \cite{5p, three}.  However, we have not yet a formal proof of the conjecture. For that we need to prove that a  system evolving without dissipation and vorticity cannot radiate gravitational radiation, ie. we have to show that during the evolution regime, after leaving the dynamic  equilibrium, $H_1=H_2=0$ all along the evolution.

Thus we consider a system which during its evolution satisfies the conditions

\begin{equation}
q_I=q_{II}=0\,\,\Rightarrow \,\, \Omega=0\,\,\Rightarrow \,\, \sigma_I=\sigma_{II}=\sigma.\label{con1}
\end{equation} 

\noindent Then from (\ref{A1}) we obtain

\begin{equation}
\frac{(2\Theta-\sigma)^\prime}{3}=\frac{\sigma C^\prime}{C},\label{el1}
\end{equation}

\noindent whereas (\ref{A2}) reads

\begin{equation}
\frac{(2\Theta-\sigma)_{,\theta}}{3}=\frac{\sigma C_{,\theta}}{C}.\label{el2}
\end{equation}

\noindent Also, (\ref{A3}), (\ref{A4}) and (\ref{A5}) become

\begin{equation}
H_1=-\frac{(\sigma C)_{,\theta}}{2rBC}\label{A11},
\end{equation}

\begin{equation}
H_2=\frac{(\sigma C)^\prime}{2BC}\label{A12},
\end{equation}

\noindent and
\begin{equation}
\frac{H_1^\prime}{B}+\frac{H_{2,\theta}}{Br}+\frac{H_1}{B}\left[ \frac{2C^\prime}{C} + \frac{(Br)^\prime}{Br} \right]+\frac{H_2}{Br}\left[\frac{2C_{,\theta}}{C} + \frac{(Br)_{,\theta}}{Br} \right]=0,\label{A19}
\end{equation}
respectively.

\noindent Using  (\ref{A11}) and (\ref{A12})  in  (\ref{A19}) we obtain
\begin{eqnarray}
&&H_1 C^\prime +H_2 \frac{C_{,\theta}}{r} =0\label{A191}\\
&&\frac{H_1^\prime}{B}+\frac{H_{2,\theta}}{Br}+\frac{H_1}{B}\frac{(Br)^\prime}{Br} +\frac{H_2}{Br} \frac{(Br)_{,\theta}}{Br} =0\label{A192},
\end{eqnarray}
from which is obvious  that the vanishing of either one of the scalars ($H_1$ or $H_2$) implies the vanishing of the other.

Finally let us notice that using (\ref{el1}) and (\ref{el2}) in (\ref{A11}) and (\ref{A12}) we may write

\begin{equation}
H_1=-\frac{1}{rB}\left(\frac{\dot B}{AB}\right)_{,\theta}\label{A11b},
\end{equation}

\begin{equation}
H_2=\frac{1}{B}\left(\frac{\dot B}{AB}\right)^\prime \label{A12b}.
\end{equation}
 
Let us now proceed to the second part of the proof. 

We start from an initially static  situation, meaning that at $t=0$,  we have $\dot A=\dot B=\dot C=\sigma=\Theta=H_1=H_2=0$. Besides, conditions (\ref{con1}) are satisfied for all $t$.

Let us take the first time derivatives of (\ref{A11b}), (\ref{A12b}), (\ref{A1}) and (\ref{A2}) evaluated at $t=0$, we obtain respectively 

\begin{equation}
 \dot H_1=-\frac{1}{rB}\left(\frac {\ddot B}{AB}\right)_{,\theta}\label{A11bb},
\end{equation}

\begin{equation}
\dot H_2=\frac{1}{B}\left(\frac{\ddot B}{AB}\right)^\prime \label{A12bb},
\end{equation}

\begin{equation}
8\pi\dot q_IB=\left[\frac{1}{A}\left(\frac{\ddot B}{B}+\frac{\ddot C}{C}\right) \right]^\prime-\left(\frac{\ddot B}{AB}-\frac{\ddot C}{AC}\right)\frac{C^\prime}{C}=0,
\label{p1}
\end{equation}

and 
\begin{equation}
8\pi\dot q_{II}Br=\left[\frac{1}{A}\left(\frac{\ddot B}{B}+\frac{\ddot C}{C}\right) \right]_{,\theta}-\left(\frac{\ddot B}{AB}-\frac{\ddot C}{AC}\right)\frac{C_{,\theta}}{C}=0.
\label{p2}
\end{equation}

From regularity conditions (\ref{nuev1}) and (\ref{nuev3}) at $r\approx0$, and from the fact that $A, B ,C$ and their derivatives are regular at $r\approx 0$ we may write at $r\approx 0$, using (\ref{p1})
\begin{equation}
W\equiv \frac{\ddot B}{AB}-\frac{\ddot C}{AC} \approx r\approx 0.
\label{p3}
\end{equation}

Taking successive  $r$-derivatives of (\ref{p1}) it is a simple matter to check that all $r$-derivatives (of any order) of $W$ vanish at $r\approx 0$, implying that $W=0$ for all values of $r$ within the fluid distribution.

Thus at $t\approx 0$ we have
\begin{equation}
\frac{\ddot B}{AB}-\frac{\ddot C}{AC}=0,
\label{p3b}
\end{equation}
for all values of $r$ within the fluid distribution.

Feeding back (\ref{p3b}) into (\ref{p2}), we obtain
\begin{equation}
\left(\frac{\ddot B}{AB}\right)_{,\theta}=\left(\frac{\ddot C}{AC}\right)_{,\theta}=0,
\label{p4b}
\end{equation}
which combined with (\ref{A11bb}) produces $\dot H_1=0$, and by virtue of (\ref{A191}), $\dot H_2=0$ as well.

Next, feeding back (\ref{p3}) into (\ref{p1}) produces 

\begin{equation}
\left(\frac{\ddot B}{AB}\right)^\prime=\left(\frac{\ddot C}{AC}\right)^\prime=0,
\label{p5b}
\end{equation}
implying because $A$ and $B$ and $C$ are independent functions, that the only admissible solution to (\ref{p5b}) (which is an identity)  is $A^\prime=0$, and $B$ and $C$ are separable functions.

The next step consists in proving  that time derivatives of any order of $H_1$ and $H_2$ evaluated at $t\approx 0$ also vanish, i.e.
\begin{equation}
\stackrel{(m)}{H_1}=\stackrel{(m)}{H_0}=0 \quad(for\; any \;  m\geq 1), {where} \stackrel{(m)}{X} \equiv \frac{\partial^m X}{\partial t^m},
\label{p9}
\end{equation}
this would imply that $H_1=H_2=0$ for any $t$.

For doing that we shall retrace the same steps above, using (\ref{con1}) and the results obtained so far.

Thus, taking the $m$-time derivative (with $m\geq1$) of (\ref{A1}) and (\ref{A2}) we may write

\begin{equation}
\frac{ \stackrel{(m+1)}B}{AB}=\frac{\stackrel{(m+1)}C}{AC}.
\label{p10}
\end{equation}

Also taking the $m$-time derivative of (\ref{A11b}) and (\ref{A12b}) produces

\begin{equation}
 \stackrel{(m)}{H_1}=-\frac{1}{Br}\left[\frac{\stackrel{(m+1)}B}{AB}\right]_{,\theta},
\label{p11}
\end{equation}

\begin{equation}
 \stackrel{(m)}{H_2}=\frac{1}{B}\left[\frac{\stackrel{(m+1)}B}{AB}\right]^\prime.
\label{p11}
\end{equation}
Using   the separability of $B$ and the fact that $A^\prime=0$, in (\ref{p11}) it follows at once that 
\begin{equation}
 \stackrel{(m)}{H_2}=0,
\label{p12}
\end{equation}
which by virtue of (\ref{A191}) implies  $\stackrel{(m)}{H_1}=0$.

Thus if the system is initially static and evolves without vorticity and without dissipation then all time derivatives of any order of $H_1$ and $H_2$ vanish for all values of $r$, implying that  $H_1=H_2=0$ at all times. 

There is yet another, perhaps more simple, way to prove  the above mentioned statement. Indeed, it is a simple matter to see that the $m$-time derivative of $\sigma$ evaluated at $t=0$, for any $m\geq 1$ reads
\begin{equation}
\stackrel{(m)}\sigma=\frac{1}{B}\left[\frac{\stackrel{(m+1)}B}{AB}-\frac{\stackrel{(m+1)}C}{AC}\right],
\label{p13}
\end{equation}
implying because  of (\ref{p10}) that time derivatives of $\sigma$ of any order vanish at $t=0$, implying in its turn that the fluid is shear--free. But as  shown in \cite{sf},   for a shear-free fluid  (not necessarily perfect fluid), the necessary and sufficient condition to be irrotational is that the Weyl tensor be purely electric; this  generalizes a result by Barnes \cite{b1,b2} and Glass \cite{glass}. Besides, it is worth noticing that $H_1=H_2=0$ implies that the fluid is also geodesic $A^\prime=A_{,\theta}=0$.

Thus  according to (\ref{SPP}) the two components of the super-Poynting vector vanish, meaning that no gravitational radiation is produced during the evolution  of the system.

\section{Conclusions}
The purpose of this work was to prove  the correctness of the Bondi conjecture about the irreversibility associated with gravitational radiation.  In other words, a reversible flow ($q=0$)  implies no gravitational radiation. We proved that, by showing that the absence of dissipative flux during the evolution (fulfillment of Tolman conditions), implies that  magnetic parts of the Weyl tensor vanish, thereby implying the vanishing of the super-Poynting vector.

At this point, it is worth stressing the fact that a reversible flow, e.g. a perfect isotropic fluid, does not necessarily imply non--crossing of flow lines (geodesic), unless we assume that pressure gradients vanish. Indeed, for a perfect (isotropic and non-dissipative)  fluid the equation of motion reads
\begin{equation}
(\mu + P) a^\alpha = h^{\alpha\nu} P_{,\nu},
\label{em}
\end{equation}
from where is clear that the geodesic condition ($a^\alpha=0$), automatically implies vanishing of pressure gradients. In this latter case if the fluid is bounded, and we impose matching conditions on the boundary surface,  then  the pressure vanishes and we have geodesic dust.

As a byproduct of our proof it appears that the vorticity of the fluid also vanishes under the condition mentioned above, bringing out, on the one hand the link of vorticity with dissipative processes already established in \cite{82}--\cite{75}, and on the other hand the link between vorticity and gravitational radiation discussed in \cite{5p, three} (and references therein).

 It is worth mentioning that the fluid configuration emerging from our restrictions, not only is non--radiative (gravitationally), shear--free, non--dissipative and vorticity free, but is also geodesic, as a consequence of which the Tolman conditions imply a homogeneous temperature.  These kind of solutions have been investigated in detail in \cite{c1,c2}. Such solutions are in general non--conformally flat,  whose electric Weyl tensor (\ref{E'}) being  defined through the  scalars
\begin{widetext}
\begin{eqnarray}
{\cal E}_I={\cal E}_I(0)exp[-\frac{2}{3}\int{\Theta dt}] ,\qquad {\cal E}_{II}={\cal E}_{II}(0)exp[-\frac{2}{3}\int{\Theta dt}],\qquad {\cal E}_{KL}={\cal E}_{KL}(0)exp[-\frac{2}{3}\int{\Theta dt}],\label{integra}
\label{int}
\end{eqnarray}
\end{widetext}
with $\Theta=\Theta(t)$.

Also, these models are characterized by the vanishing of the trace--free part of the  tensor $Y_{\alpha \beta}$ (see \cite{sf} for details), i.e.

\begin{equation}
Y_I=Y_{KL}=Y_{II}=0.\label{losY}
\end{equation}

Parenthetically, these three scalars haven been proposed to describe the degree of complexity  of a fluid distribution \cite{com1,com2}. Thus according to the criterium assumed in these references, the resulting models are the simplest among those belonging to the family of space--times described by (\ref{1b}).

Finally, we would like to conclude with two  remarks
\begin{itemize}
 
 \item The fact that the emission of gravitational radiation  requires  the presence of  dissipative flux within the source  to account by the irreversibility of the process, implies that any detected burst of gravitational waves should be accompanied by a burst of thermal radiation, which in principle could be observed too. 
\item  An alternative way of proving the Bondi's conjecture could be provided by assuming  a perfect fluid (so that $T = (mu+p)VV+pg)$ with an equation of state $p
= p(\mu, s)$ where $\mu$ is the energy density and $s$ the specific
entropy.  Then,  from the generalized Gibbs equation and the Bianchi identities, it follows that (see for example \cite{ul})
\begin{widetext}
\begin{eqnarray}
T S^{\alpha}_{;\alpha} =  
- q^{\alpha} \left[ h^\mu_{\alpha} (\ln{T })_{,\mu} + 
V_{\alpha;\mu} V^\mu
 + \beta_{1} q_{\alpha;\mu} V^\mu+
\frac{T}{2} \left(
\frac{\beta_{1}}{T}V^{\mu}\right)_{;\mu}q_{\alpha}\right],
\label{diventropia}
\end{eqnarray}
\end{widetext}
where $S^\alpha$ is the entropy four--current and $\beta_1=\frac{\tau}{\kappa T} $.
If we assume from the beginning that the matter content of the source is a perfect fluid (no heat flux vector) then $S^{\alpha}_{;\alpha}=0$, implying that the entropy is constant and Tolman conditions are satisfied. From this point,  there are different ways to prove that no gravitational radiation is produced, one of which is the one   we have chosen in this manuscript, though it is not the only one. 
\end{itemize}

\section{Acknowledgements} This work was partially supported by the Grant PID2021-122938NB-I00 funded by MCIN/AEI/ 10.13039/501100011 033 and by ``ERDF A way of making Europe''.
\appendix 
\section{Summary of scalar variables and equations}
The required equations for our proof are given explicitly in \cite{1}. Here for self--consistency we present a brief summary of them,  including only those equations explicitely required for our proof.  The reader is referred to \cite{1} for details of calculations.

The anisotropic tensor  may be  expressed in the form 
\begin{widetext}
\begin{eqnarray}
\Pi_{\alpha \beta}=\frac{1}{3}(2\Pi_I+\Pi_{II})\left(K_\alpha K_\beta-\frac{h_{\alpha
\beta}}{3}\right)+\frac{1}{3}(2\Pi _{II}+\Pi_I)\left(L_\alpha L_\beta-\frac{h_{\alpha
\beta}}{3}\right)+2\Pi _{KL}K_{(\alpha}L_{\beta)} \label{6bb},
\end{eqnarray}
\end{widetext}
with

 $h_{\mu \nu}=g_{\mu\nu}+V_\nu V_\mu$,

\begin{eqnarray}
 \Pi _{KL}=K^\alpha L^\beta T_{\alpha \beta} 
, \quad \label{7P}
\end{eqnarray}

\begin{equation}
\Pi_I=(2K^{\alpha} K^{\beta} -L^{\alpha} L^{\beta}-S^{\alpha} S^{\beta}) T_{\alpha \beta},
\label{2n}
\end{equation}
\begin{equation}
\Pi_{II}=(2L^{\alpha} L^{\beta} -S^{\alpha} S^{\beta}-K^{\alpha} K^{\beta}) T_{\alpha \beta}.
\label{2nbis}
\end{equation}

The heat flux vector may be written as 
\begin{equation}
q_\mu=q_IK_\mu+q_{II} L_\mu,
\label{qn1}
\end{equation}
or, in coordinate components

\begin{equation}
q^\mu=\left(\frac{q_{II} G}{A \sqrt{A^2B^2r^2+G^2}}, \frac{q_I}{B}, \frac{Aq_{II}}{\sqrt{A^2B^2r^2+G^2}}, 0\right),\label{q}
\end{equation}
\begin{equation}
 q_\mu=\left(0, B q_I, \frac{\sqrt{A^2B^2r^2+G^2}q_{II}}{A}, 0\right).
\label{qn}
\end{equation}
Of course, all the above quantities depend,  in general, on $t, r, \theta$.

The kinematical variables (four--acceleration,  expansion scalar, shear tensor, and  vorticity), are

\begin{eqnarray}
a_\alpha&=&V^\beta V_{\alpha;\beta}=a_I K_\alpha+a_{II}L_\alpha\nonumber\\
&=&\left(0, \frac {A^\prime }{A },\frac{G}{A^2}\left[-\frac {\dot A}{A}+\frac {\dot G}{G}\right]+\frac {A_{,\theta}} {A},0\right),
\label{acc}
\end{eqnarray}
\begin{eqnarray}
\Theta&=&V^\alpha_{;\alpha}\nonumber\\
&=&\frac{AB^2}{r^2A^2B^2+G^2}\,\left[r^2\left(2\frac{\dot B}{B}+\frac{\dot C}{C}\right)\right.\nonumber\\
&&+\left.\frac{G^2}{A^2B^2}\left(\frac{\dot B}{B}-\frac{\dot A}{A}+\frac{\dot G}{G}+\frac{\dot C}{C}\right)\right],
\label{theta}
\end{eqnarray}

\begin{eqnarray}
\sigma _{\alpha \beta}=\frac{1}{3}(2\sigma _I+\sigma_{II}) \left(K_\alpha
K_\beta-\frac{1}{3}h_{\alpha \beta}\right)\nonumber \\+\frac{1}{3}(2\sigma _{II}+\sigma_I) \left(L_\alpha
L_\beta-\frac{1}{3}h_{\alpha \beta}\right),\label{sigmaT}
\end{eqnarray}

where
\begin{eqnarray}
2\sigma _I+\sigma_{II}&=&\frac{3}{A}\left(\frac{\dot B}{B}-\frac{\dot C}{C}\right), \label{sigmasI}
\end{eqnarray}
\begin{eqnarray}
2\sigma _{II}+\sigma_I&=&\frac{3}{A^2B^2r^2+G^2}\,\left[AB^2r^2\left(\frac{\dot B}{B}-\frac{\dot C}{C}\right)\right.\nonumber\\
&
&\left.+\frac{G^2}{A}\left(-\frac{\dot A}{A}+\frac{\dot G}{G}-\frac{\dot C}{C}\right)\right] \label{sigmas},
\end{eqnarray}
in the above  dots and primes denote derivatives with respect to $t$ and $r$ respectively. 

Finally, for the vorticity vector defined as
\begin{equation}
\omega_\alpha=\frac{1}{2}\,\eta_{\alpha\beta\mu\nu}\,V^{\beta;\mu}\,V^\nu=\frac{1}{2}\,\eta_{\alpha\beta\mu\nu}\,\Omega
^{\beta\mu}\,V^\nu,\label{vomega}
\end{equation}
where $\Omega_{\alpha\beta}=V_{[\alpha;\beta]}+a_{[\alpha}
V_{\beta]}$ and $\eta_{\alpha\beta\mu\nu}$ denote the vorticity tensor and the Levi-Civita tensor respectively,  we find a single component different from zero,  producing

\begin{equation}
\Omega_{\alpha\beta}=\Omega (L_\alpha K_\beta -L_\beta
K_{\alpha}),\label{omegaT}
\end{equation}
and
\begin{equation}
\omega _\alpha =-\Omega S_\alpha,
\end{equation}
with the scalar function $\Omega$ given by
\begin{equation}
\Omega =\frac{G(\frac{G^\prime}{G}-\frac{2A^\prime}{A})}{2B\sqrt{A^2B^2r^2+G^2}}.
\label{no}
\end{equation}

Now, from the regularity conditions, necessary to ensure elementary flatness in the vicinity of  the axis of symmetry, and in particular at the center (see \cite{1n}, \cite{2n}, \cite{3n}), we should require  that as $r\approx 0$
\begin{equation}
\Omega=\sum_{n \geq1}\Omega^{(n)}(t,\theta) r^{n},
\label{sum1}
\end{equation}
implying, because of (\ref{no}) that in the neighborhood of the center
\begin{equation}
 G=\sum_{n\geq 3} G^{(n)}(t,\theta) r^{n}.
\label{sum1}
\end{equation}

Also, for the length of an orbit at $t, \theta$ constant, to be $2\pi r$, close to the origin (elementary flatness), we may write, as $r\rightarrow 0$, 
\begin{equation}
C\approx r\gamma(t,\theta), 
\label{nuev1}
\end{equation}
implying
\begin{equation}
C^\prime \approx \gamma(t,\theta), \;\; C_{,\theta}\approx r\gamma_{,\theta},
\label{nuev3}
\end{equation}
where $\gamma(t,\theta)$ is an arbitrary function of its arguments, which as appears evident from the elementary flatness condition, cannot vanish anywhere within the fluid distribution.

Observe that from (\ref{no}) and regularity conditions at the centre, it follows that: $G=0\Leftrightarrow \Omega=0$.

Next, for  the electric ( $E_{\alpha\beta}$)
and magnetic  ($H_{\alpha\beta}$) parts of the Weyl tensor $C_{\alpha \beta
\gamma\delta}$, we have
\begin{eqnarray}
E_{\alpha \beta}&=&C_{\alpha\nu\beta\delta}V^\nu V^\delta,\nonumber\\
H_{\alpha\beta}&=&\frac{1}{2}\eta_{\alpha \nu \epsilon
\rho}C^{\quad \epsilon\rho}_{\beta \delta}V^\nu
V^\delta\,.\label{EH}
\end{eqnarray}

The electric part of the Weyl tensor has only three independent non-vanishing components, whereas only two components define the magnetic part. Thus  we may also write

\begin{widetext}
\begin{equation}
E_{\alpha\beta}=\frac{1}{3}(2\mathcal{E}_I+\mathcal{E}_{II}) \left(K_\alpha
K_\beta-\frac{1}{3}h_{\alpha \beta}\right) +\frac{1}{3}(2\mathcal{E}_{II}+\mathcal{E}_{I}) \left(L_\alpha
L_\beta-\frac{1}{3}h_{\alpha \beta}\right)+\mathcal{E}_{KL} (K_\alpha
L_\beta+K_\beta L_\alpha), \label{E'}
\end{equation}
\end{widetext}
\noindent

and
\begin{equation}
H_{\alpha\beta}=H_1(S_\alpha K_\beta+S_\beta
K_\alpha)+H_2(S_\alpha L_\beta+S_\beta L_\alpha)\label{H'}.
\end{equation}

The orthogonal splitting of the Riemann tensor is  carried out by
means of three tensors $Y_{\alpha\beta}$, $X_{\alpha\beta}$ and
$Z_{\alpha\beta}$ defined as

\begin{equation}
Y_{\alpha \beta}=R_{\alpha \nu \beta \delta}V^\nu V^\delta,
\label{Y}
\end{equation}
\begin{equation}
X_{\alpha \beta}=\frac{1}{2}\eta_{\alpha\nu}^{\quad \epsilon
\rho}R^\star_{\epsilon \rho \beta \delta}V^\nu V^\delta,\label{X}
\end{equation}
and
\begin{equation}
Z_{\alpha\beta}=\frac{1}{2}\epsilon_{\alpha \epsilon \rho}R^{\quad
\epsilon\rho}_{ \delta \beta} V^\delta,\label{Z}
\end{equation}
 where $R^\star _{\alpha \beta \nu
\delta}=\frac{1}{2}\eta_{\epsilon\rho\nu\delta}R_{\alpha
\beta}^{\quad \epsilon \rho}$, and $\epsilon _{\alpha \beta \rho}=\eta_{\nu
\alpha \beta \rho}V^\nu$.

The three tensors above may be expressed through the following scalars functions
\begin{eqnarray}
Y_T=4\pi(\mu+3P), \label{ortc1}\\
Y_I=\mathcal{E}_I-4\pi \Pi_I, \label{ortc2}\\
Y_{II}=\mathcal{E}_{II}-4\pi \Pi_{II}, \label{YY}\\
Y_{KL}=\mathcal{E}_{KL}-4\pi \Pi_{KL},\label{KL}
\end{eqnarray}

\begin{eqnarray}
X_T=8\pi \mu, \label{ortc1x}\\
X_I=-\mathcal{E}_I-4\pi \Pi_I, \label{ortc2x}\\
X_{II}=-\mathcal{E}_{II}-4\pi \Pi_{II}, \label{YYx}\\
X_{KL}=-\mathcal{E}_{KL}-4\pi \Pi_{KL},\label{KLx}
\end{eqnarray}

\begin{widetext}
\begin{equation}
Z_I=(H_1-4\pi q_{II});\quad Z_{II}=(H_1+4\pi  q_{II}); \quad Z_{III}=(H_2-4\pi q_I); \quad  Z_{IV}=(H_2+4\pi q_I). \label{Z2}
\end{equation}
\end{widetext}

In the above, the scalars $Y_T, X_T$ define the trace of (\ref{Y}) and (\ref{X}), respectively, whereas the scalars $Y_I, Y_{II}, Y_{KL}, X_I, X_{II}, X_{KL}$ define the trace--free part of (\ref{Y}) and (\ref{X}).

The super--Poynting
vector defined by
\begin{equation}
P_\alpha = \epsilon_{\alpha \beta \gamma}\left(Y^\gamma_\delta
Z^{\beta \delta} - X^\gamma_\delta Z^{\delta\beta}\right),
\label{SPdef}
\end{equation}
 can be written as:
\begin{equation}
 P_\alpha=P_I K_\alpha+P_{II} L_\alpha,\label{SP}
\end{equation}

with
\begin{widetext}
\begin{eqnarray}
P_I &=
&\frac{H_2}{3}(2Y_{II}+Y_I-2X_{II}-X_I)+H_1(Y_{KL}-X_{KL}) + \frac{4\pi q_I}{3}\left[2Y_T+2 X_T-X_I-Y_I\right] \nonumber \\
&-& 4\pi q_{II}(X_{KL} +Y_{KL}),\nonumber
\\
P_{II}&=&\frac{H_1}{3}(2X_{I}+X_{II}-Y_{II}-2Y_I)+H_2(X_{KL}-Y_{KL})-4\pi q_I(Y_{KL}+X_{KL}) \nonumber \\
&+& \frac{4\pi q_{II}}{3}\left[2Y_T+2X_T-X_{II}-Y_{II}\right]. \label{SPP}
\end{eqnarray}
\end{widetext}

As mentioned before, in the theory of  the super--Poynting vector, a state of gravitational radiation is associated to a  non--vanishing component of the latter (see \cite{11p, 12p, 14p}). Therefore we shall verify the absence of gravitational radiation if  the two components of the super--Poynting vector vanish.

From the  Ricci identities for the vector
$V_\alpha$, and the Bianchi identities
the following set of equations are obtained by contracting with different vectors of the tetrad (see \cite{1} for details).  These are:

An evolution equation for $\Omega$ (Eq.(B5 ) in \cite{1})
\begin{equation}
\Omega _{,\delta}V^\delta +\frac{1}{3}(2\Theta+\sigma _I+\sigma _{II})\Omega +K^{[\mu}L^{\nu]}a_{\mu;\nu}=0,\label{esc51KL}
\end{equation}
where
\begin{widetext}
\begin{eqnarray}
\left(K^\mu L^\nu-L^\mu K^\nu\right) a_{\nu;\mu} = -\frac{A}{\sqrt{A^2 B^2 r^2 + G^2}}\left[a_{I,\theta}+\frac{G}{A^2}\dot a_{I}+a_{I}\left(\frac{B_{,\theta}}{B}+\frac{G}{A^2} \frac{\dot B}{B}\right)\right]\nonumber\\
+\frac{1}{B}\left[a_{II}^{\prime}+\frac{a_{II}}{A^2 B^2 r^2 + G^2}\left(A^2B^2r^2 \frac{(Br)^{\prime}}{Br}+GG^{\prime}-G^2\frac{A^{\prime}}{A}\right)\right],\nonumber\\
\label{T7}
\end{eqnarray}
\end{widetext}
with
\begin{equation}
a_{II}=\frac{A a_2}{\sqrt{A^2B^2r^2+G^2}},\qquad  a_I=\frac{a_1}{B}.
\label{ace1}
\end{equation}

Two equations relating $q_I$ and $ q_{II}$ with the kinematical variables (Eqs. (B6) and (B7) in \cite{1})
\begin{widetext}
\begin{eqnarray}
\frac{2}{3B}\Theta _{,r}-\Omega _{;\mu}L^\mu+\Omega (L_{\beta ;\mu}K^\mu K^\beta-L^\mu _{;\mu})+\frac{1}{3}\sigma _I a_I-\Omega a_{II}-\frac{1}{3}\sigma_{I;\mu}K^\mu\nonumber
\\
-\frac{1}{3}(2\sigma _I+\sigma _{II})(K^\mu _{;\mu}-\frac{a_I}{3})-\frac{1}{3}(2\sigma _{II}+\sigma _I)(L_{\beta ;\mu}L^\mu K^\beta-\frac{a_I}{3})=8\pi
q_I,\label{A1}
\end{eqnarray}
\end{widetext}
\begin{widetext}
\begin{eqnarray}
\frac{1}{3\sqrt{A^2B^2r^2+G^2}}\left(\frac{2G}{A}
\Theta_{,t}+2A\Theta _{,\theta}\right )+\frac{a_{II} \sigma _{II}}{3}+\Omega _{;\mu}K^\mu+\Omega (K^\mu _{;\mu}+L^\mu K^\beta L_{\beta;\mu})+\Omega a_I-\frac{1}{3}\sigma_{II;\mu}L^\mu \nonumber\\
+\frac{1}{3}(2\sigma _I+\sigma_{II})(L_{\beta;\mu} K^\beta K^\mu+\frac{a_{II}}{3})-\frac{1}{3}(2\sigma _{II}+\sigma _I)(L^\mu _{;\mu}-\frac{a_{II}}{3})=8\pi q_{II}.
\label{A2}
\end{eqnarray}
\end{widetext}

Two equations  relating the two scalars defining the magnetic part of the Weyl tensor with the kinematical variables (Eqs. (B8) and (B9) in \cite{1})
\begin{equation}
-\Omega a_I-\frac{1}{2}(K^\mu S_\nu+S^\mu K_\nu)(\sigma _{\mu\delta}+\Omega _{\mu\delta})_{;\gamma}\epsilon ^{\nu\gamma\delta}=H_1,\label{A3}
\end{equation}
\begin{equation}
-\Omega a_{II}-\frac{1}{2}(L^\mu S_\nu+S^\mu L_\nu)(\sigma _{\mu\delta}+\Omega _{\mu\delta})_{;\gamma}\epsilon ^{\nu\gamma\delta}=H_2.\label{A4}
\end{equation}

Also, from one of equations derived from the Bianchi identities, one obtains (Eq. (B16) in \cite{1})
\begin{widetext}
\begin{eqnarray}
-\frac{1}{3}X_{KL}(\sigma _{II}-\sigma
_I)+a_IH_1+a_{II} H_2 -H_{1,\delta}K^\delta-H_{2,\delta}L^\delta-H_1(K^\delta _{;\delta}+K^\nu _{;\delta}S^\delta S_\nu)\nonumber
 \\
-H_2(L^\delta _{;\delta}+S^\delta S_\nu L^\nu _{;\delta}) =\left \{8\pi [\mu +P-\frac{1}{3}(\Pi _I+\Pi _{II})]-Y_I-Y_{II}\right \}\Omega
 -\frac{4\pi A (q_IB)_{,\theta}}{B\sqrt{A^2B^2r^2+G^2}}\nonumber \\
 +\frac{4 \pi A}{B \sqrt{A^2B^2r^2+G^2}} \left[\frac{q_{II} \sqrt{(A^2B^2r^2+G^2)}}{A}\right]_{,r}.
  \label{A5}
 \end{eqnarray}
\end{widetext}

\end{document}